\documentclass[aps,prl,notitlepage,10pt,twocolumn,superscriptaddress]{revtex4-1}
\usepackage[utf8]{inputenc}
\usepackage{geometry}
\usepackage{xcolor}

\usepackage{hyperref}
\usepackage{graphicx}
\usepackage{amsmath}
\usepackage{wasysym}
\usepackage{stmaryrd}
\usepackage{subfigure}
\usepackage{listings}
\usepackage{color}
\usepackage[T1]{fontenc}
\usepackage{tabularx}

\begin{document}
\title{Not-So-Glass-Like Caging and Fluctuations of an Active Matter Model}
\author{Mingyuan Zheng}
\email[Author to whom correspondence should be addressed. Electronic mail: ]{mingyuan.zheng@duke.edu}
\affiliation{Department of Chemistry, Duke University, Durham, North Carolina 27708, United States}
\author{Dmytro Khomenko}
\affiliation{Dipartimento di Fisica, Sapienza Universit\`a di Roma, Piazzale Aldo Moro 5, 00185 Rome, Italy}
\affiliation{NOMATEN Centre of Excellence, National Center for Nuclear Research, ul. A. Soltana 7, 05-400 Swierk/Otwock, Poland}
\author{Patrick Charbonneau}
\affiliation{Department of Chemistry, Duke University, Durham, North Carolina 27708, United States}
\affiliation{Department of Physics, Duke University, Durham, North Carolina 27708, United States}
\date{\today}
\begin{abstract}
Simple active models of matter recapitulate complex biological phenomena. The out-of-equilibrium nature of these models, however, often makes them beyond the reach of first-principle descriptions. This limitation is particularly perplexing when attempting to distinguish between different glass-forming mechanisms. We here consider a minimal active system in various spatial dimensions to identify the processes underlying their sluggish dynamics. Activity is found to markedly impact cage escape processes and critical fluctuations associated with exploring lower-dimensional caging features. 

\end{abstract}
\date{\today}
\maketitle

\paragraph{Introduction -- }
Remarkably collective processes are observed in both macroorganisms, such as bird flocks and insect swarms~\cite{ramaswamy2010mechanics,marchetti2013hydrodynamics,cavagna2014bird}, and microorganisms, such as swimming bacteria and motile cells~\cite{elgeti2015physics,polin2009chlamydomonas,ghosh2013self,maass2016swimming}. Equally remarkable is that the essence of these phenomena can be recapitulated by schematic models of active, or \textit{self-propelled}, particles whose motion persists along a given direction~\cite{romanczuk2012active,bechinger2016active,zottl2016emergent,nandi2018random,mandal2020multiple}. Even these simplified models, however, remain challenging to describe from first principles. Hence, while the physical origin of motility-induced phase separation is now fairly clear~\cite{cates2015motility,redner2013structure,fily2014freezing,stenhammar2014phase,digregorio2018full,de2019active,paoluzzi2022motility}, the nature of dynamical arrest at higher densities remains debated~\cite{janssen2018mode,berthier2019glassy,janssen2019active}. 
Active glasses are often deemed equivalent to their passive counterparts with quantitatively enhanced (effective) diffusion~\cite{howse2007self,farage2014dynamics,flenner2016nonequilibrium} and temperature~\cite{berthier2013non,nandi2017nonequilibrium,cugliandolo2019effective}, and longer ranged velocity correlations~\cite{szamel2016theory,caprini2020hidden}. Yet some of their phenomenological features, such as their dynamical heterogeneity~\cite{paul2023dynamical}, are qualitatively distinct. The extent to which arrest in active systems is akin to that of standard liquid glass formers therefore remains unclear~\cite{berthier2017active,ni2013pushing,berthier2014nonequilibrium}. 

Descriptions from either spin-glass theory~\cite{berthier2013non}, mode-coupling theory (MCT)~\cite{janssen2018mode,farage2014dynamics,liluashvili2017mode,feng2017mode,szamel2019mode,reichert2021mode}, or dynamical mean-field theory (DMFT)~\cite{agoritsas2019out1,agoritsas2019out2,morse2021direct} offer some insight into the matter, but also suffer from severe limitations. The first is based on a loose analogy, the second on uncontrolled approximations~\cite{pihlajamaa2023unveiling}, and the third cannot be solved using existing numerical tools (other than in the dilute limit~\cite{manacorda2022gradient}). In that sense, the problem is even more challenging than that of simple glasses for which these descriptions have traditionally been used~\cite{charbonneau2017glass,parisi2020theory}. In this context, one promising approach is to employ numerical simulations to interpolate between systems of interest and the infinite-dimensional limit, $d\rightarrow\infty$, in which the mean-field description is exact. This analysis has been strikingly effective in elucidating the behavior of simple glass formers~\cite{parisi2020theory}, notably by identifying the importance of marginality around the jamming transition~\cite{charbonneau2017glass}, sorting out the algorithmic dependence of jamming~\cite{charbonneau2023jamming}, and distinguishing between perturbative mean-field-like caging~\cite{biroli2022local,folena2022equilibrium,charbonneau2024dynamics,laudicina2024} and nonperturbative instantonlike single-particle hopping~\cite{biroli2021interplay,charbonneau2014hopping,dzero2009replica}, around the dynamical arrest. 

A particularly promising system for conducting such an analysis for active glass formers is the active Brownian particle random Lorentz gas (ABP-RLG)~\cite{ten2011brownian,de2018static,zeitz2017active,sprenger2020active,janssen2019active,zeitz2017active}. Like the passive RLG -- a prototypical continuum-space model for kinetic theory~\cite{dorfman2021contemporary} and percolation~\cite{elam1984critical,meester1996continuum,rintoul1997precise,charbonneau2021high,hofling2006localization,hofling2008critical,jin2015dimensional,charbonneau2021high} -- the ABP-RLG evolves a tracer within the void space left by fixed spherical obstacles. Interestingly, the resulting interplay between dynamics and geometry has the same mean-field description as simple (hard-sphere-like) liquids in the limit $d\rightarrow\infty$~\cite{biroli2021interplay,biroli2021mean,manacorda2022gradient}, and can therefore be used to distinguish single-particle from genuinely collective phenomena in finite $d$. Prior studies of ABP-RLG have mostly focused on the percolation regime~\cite{zeitz2017active}, which--given its purely geometrical nature--is unsurprisingly unchanged from that of the standard RLG. The glasslike arrest, which is expected to be much more sensitive to activity~\cite{ni2013pushing,berthier2014nonequilibrium,liluashvili2017mode,keta2022disordered}, is much less understood. 

In this Letter, using simulations of the ABP-RLG in spatial dimensions $d=2\ldots 12$,  we find that activity results in lower-dimensional caging features being explored in the dynamically sluggish regime, thereby (i) enhancing escape processes by sidestepping instantonlike hopping, and (ii) giving rise to distinct critical fluctuations. In short, we find that activity alters the nature of the dynamical arrest from that of standard glass formers. 

\paragraph{Model and methods --}
\begin{figure*}[tbh]
\centering
\includegraphics[scale=0.44,trim={1.3cm 0.9cm 0.1cm 0.8cm},clip]{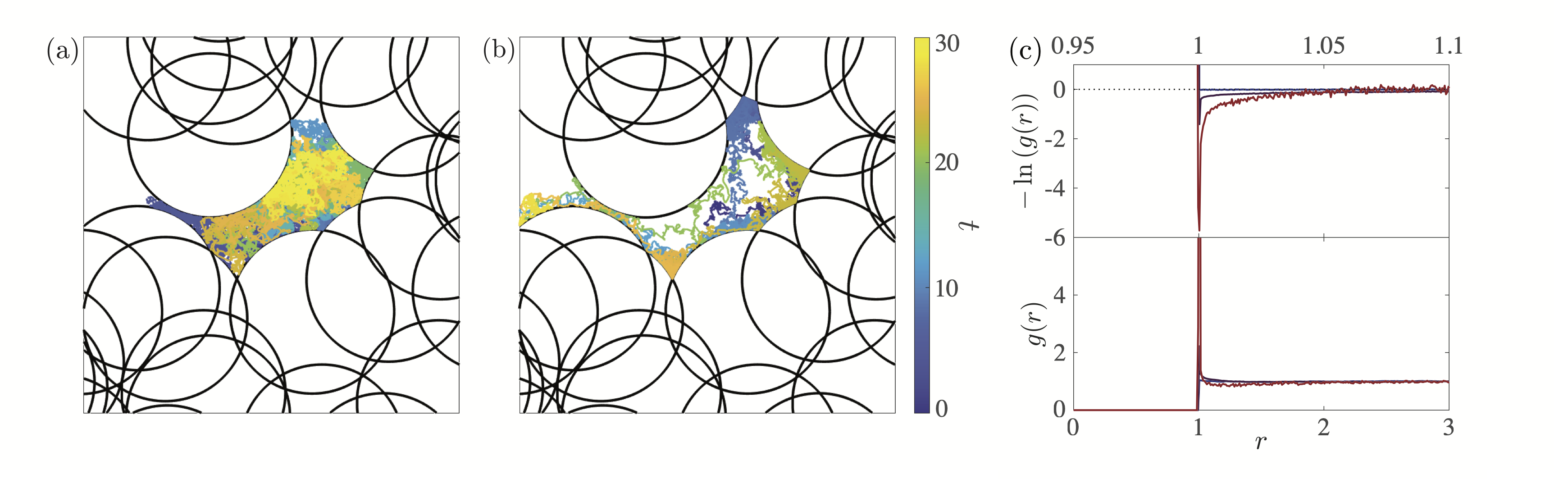}
\caption{Sample $d=2$ trajectory up to $t=30$ of a tracer with (a) $\widehat{\mathrm{Pe}} = 0$ (BP-RLG) and (b) $10$ (ABP-RLG) within the same fixed obstacles at $\widehat{\varphi} = 1.5$ (drawn to unit diameter). The passive tracer uniformly samples void space and rarely escapes the entropic bottleneck; the active tracer preferentially explores the surface of the obstacles and easily traverses the narrow path out. Activity alters the relaxation process. (c) Steady-state (bottom) radial distribution function of obstacles around the tracer, $g(r)$, and (top) corresponding  effective potential  for $\widehat{\mathrm{Pe}} = 0, 1, 10$ (blue to red lines). The contact peak markedly grows with activity up to several $k_\mathrm{B}T$, indicating a stronger effective attraction between the tracer and the obstacles.}
\label{fig:traj}
\end{figure*}
The RLG model follows the dynamics of a spherical \textit{tracer} of radius $R$ within the void space left by fixed spherical obstacles also of radius $R$, distributed uniformly at random at a (dimensionally rescaled~\cite{biroli2021interplay,biroli2022local,zottl2016emergent}) volume fraction $\widehat{\varphi} = \rho R^d V_d/d$ for number density $\rho$, where $V_d$ is the volume of $d$-dimensional unit sphere. For the ABP-RLG, the tracer is assigned a constant intrinsic velocity $v_0$ along the direction of a unitary vector $\mathbf{e}(t)$~\cite{howse2007self}. The associated overdamped Langevin equations with translational $\mathbf{\xi}_t$ and rotational $\mathbf{\xi}_r$ Gaussian white noise of zero mean and unit variance,
\begin{equation}
    \langle \mathbf{\xi}_{r,t}(t)\rangle = 0, \quad
    \langle \mathbf{\xi}_{r,t}(t)\cdot\mathbf{\xi}_{r,t}(t')\rangle = \mathbf{\delta}(t-t'),
\end{equation} 
then have
\begin{align}
    \dot{\mathbf{r}}(t) &= v_0 \mathbf{e}(t) + \sqrt{2D_t}\mathbf{\xi}_t(t),\label{eq:abp_dynamics1}\\
    \dot{\mathbf{e}}(t) &= \sqrt{2D_r}\mathbf{\xi}_r(t)\otimes \mathbf{e}(t),\label{eq:abp_dynamics2}
\end{align}
where translational and rotational diffusion constants are related as $D_r = d D_t$, thus generalizing to arbitrary $d$ the natural choice made for $d=3$ systems \cite{dhont1996introduction}. Note that this particular choice of scaling, however, does not qualitatively alter the conclusions about the role of activity~\cite{SI}. Here $\otimes$ denotes the projection of $\mathbf{\xi}_r(t)$ on the tangent surface perpendicular to $\mathbf{e}(t)$. Equation~\eqref{eq:abp_dynamics2} therefore corresponds to $\mathbf{e}(t)$ evolving on the surface of a unit hypersphere. Our simulations leverage a standard event-driven Brownian dynamics scheme~\cite{foffi2005scaling,scala2007event,charbonneau2024dynamics} with the Brownian component giving rise to purely ballistic collisions, thus preventing tracer and obstacles from ever overlapping. 

Activity strength is characterized by a P\'eclet number $\mathrm{Pe}$. Prior simulations have defined this quantity as the ratio of the diffusive timescale $4R^2/D_t$ and the ballistic time scale $2R/v_0$, i.e., $\mathrm{Pe}^s = \frac{2v_0 R}{D_t}$~\cite{zottl2016emergent,zeitz2017active}, whereas MCT studies have used $\mathrm{Pe}^{\mathrm{MCT}} = \frac{v_0^2}{D_t D_r}$~\cite{liluashvili2017mode} to obtain an effective diffusion coefficient $D_{\mathrm{eff}}=D_t(1+\mathrm{Pe}^{\mathrm{MCT}})$. In order to relate the two, we here empirically define
\begin{equation}
\label{eq:P\'{e}clet_number}
    \mathrm{Pe} = \frac{v_0}{\sqrt{D_t D_r}},
\end{equation}
and more specifically the dimensionally rescaled $\widehat{\mathrm{Pe}} = \mathrm{Pe}/\sqrt{d(d-1)}$, which recovers the expression for $D_{\textrm{eff}}$ in Eq.~\eqref{eq:abp_msd_app}, and gives $\widehat\mathrm{Pe}\approx 1$ as crossover. That is, $\widehat{\mathrm{Pe}}\lesssim 1$ corresponds to (translational) diffusion dominating -- with a standard Brownian particle (BP) having $\widehat{\mathrm{Pe}}=0$ -- while $\widehat{\mathrm{Pe}}\gtrsim 1$ corresponds to active motion dominating~\cite{SI}. As illustrated in Fig.~\ref{fig:traj}, a BP tracer samples void space uniformly with entropic bottlenecks appearing in narrow passages; an ABP tracer, by contrast, mostly explores the surface of obstacles and traverses narrow passages fairly easily, but can get wedged between obstacles, as previously noted in similar systems~\cite{bechinger2016active}. The tracer can only leave a cavity formed by obstacles if its velocity vector points outward or has at least one component parallel to the tangent surface, a process mostly controlled by the rotational component of the diffusion at high $\widehat{\mathrm{Pe}}$. (This effect becomes more significant and intricate for $d \geq 3$, because determining the moving direction then requires considering multiple obstacles at once~\cite{SI}.) The resulting increase in effective attraction between the tracer and the obstacles can be gleaned from the radial distribution function, which sharply peaks at contact as $\widehat{\mathrm{Pe}}$ increases (see Fig.~\ref{fig:traj}(c)), consistent with observations from other active systems~\cite{berthier2019glassy}.

\paragraph{Results -- }
We first consider the evolution with time $t$ of the tracer mean squared displacement (MSD), $\Delta (t)=\langle |\mathbf{r}(t)-\mathbf{r}(0)|^2\rangle$. In free space~\cite{zottl2016emergent},

\begin{equation}
\label{eq:msd_abp1}
    \Delta = 2dD_t t + 2v_0^2\tau_r t - 2 v_0^2 \tau_r^2(1-e^{-\frac{t}{\tau_r}}),
\end{equation}
where $\tau_r = \frac{1}{(d-1)D_r}$ is the persistence time of a spherical particle. With appropriate dimensional rescaling,
\begin{equation}
\label{eq:msd_abp2}
    \hat{\Delta} = \hat{t} + \widehat{\mathrm{Pe}}^2(\hat{t}-1+e^{-\hat{t}}),
\end{equation}
where $\hat{t} = t/\tau_r$, and $\widehat{\Delta} = \frac{\Delta}{2dD_t\tau_r}$ is accordingly set so that a free-space BP tracer has $\widehat{\Delta} = \hat{t}$~\cite{SI}. The rescaled P\'{e}clet number then collapses the short- and long-time free-space diffusion regimes for all $d$,
\begin{equation}
\label{eq:abp_msd_app}
    \widehat{\Delta}(\hat{t}) =
\begin{cases}
    \hat{t} + \frac{1}{2} \widehat{\mathrm{Pe}}^2 \hat{t}^2  & \mbox{for $\hat{t}\ll 1$},\\
    (1+\widehat{\mathrm{Pe}}^2)\hat{t} & \mbox{for $\hat{t}\gg 1$}.
\end{cases}
\end{equation}
Put differently, the MSD of a free ABP is diffusive ($\sim t$) at very short times, ballistic ($\sim t^2$) at intermediate times, and diffusive with an enhanced effective diffusion coefficient at long times~\cite{bechinger2016active}.

\begin{figure*}[hbt]
\centering
\includegraphics[scale=0.45,trim={0.3cm 0 0.6cm 0.6cm},clip]{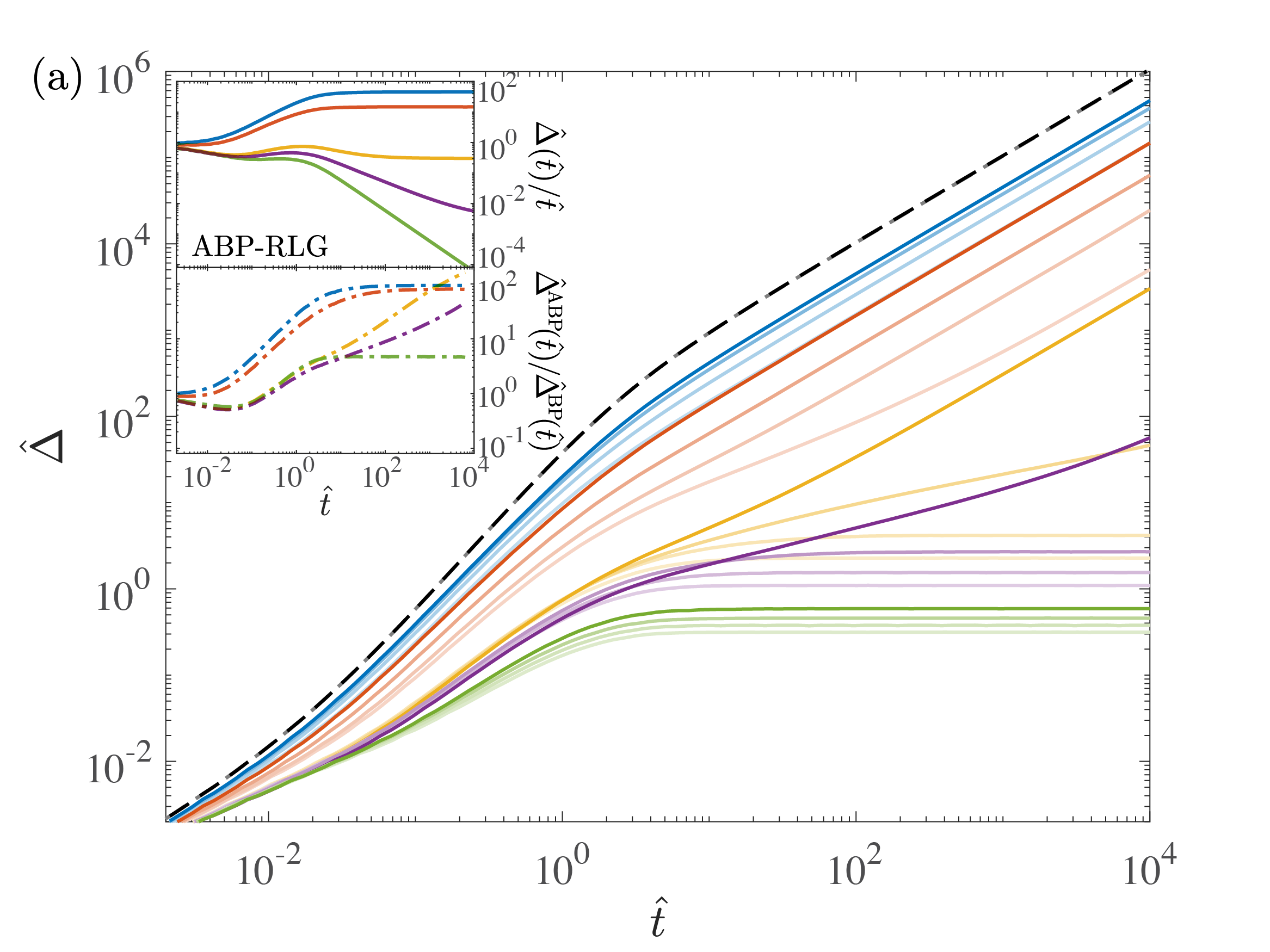}
\includegraphics[scale=0.45,trim={0.5cm 0 0.3cm 0.6cm},clip]{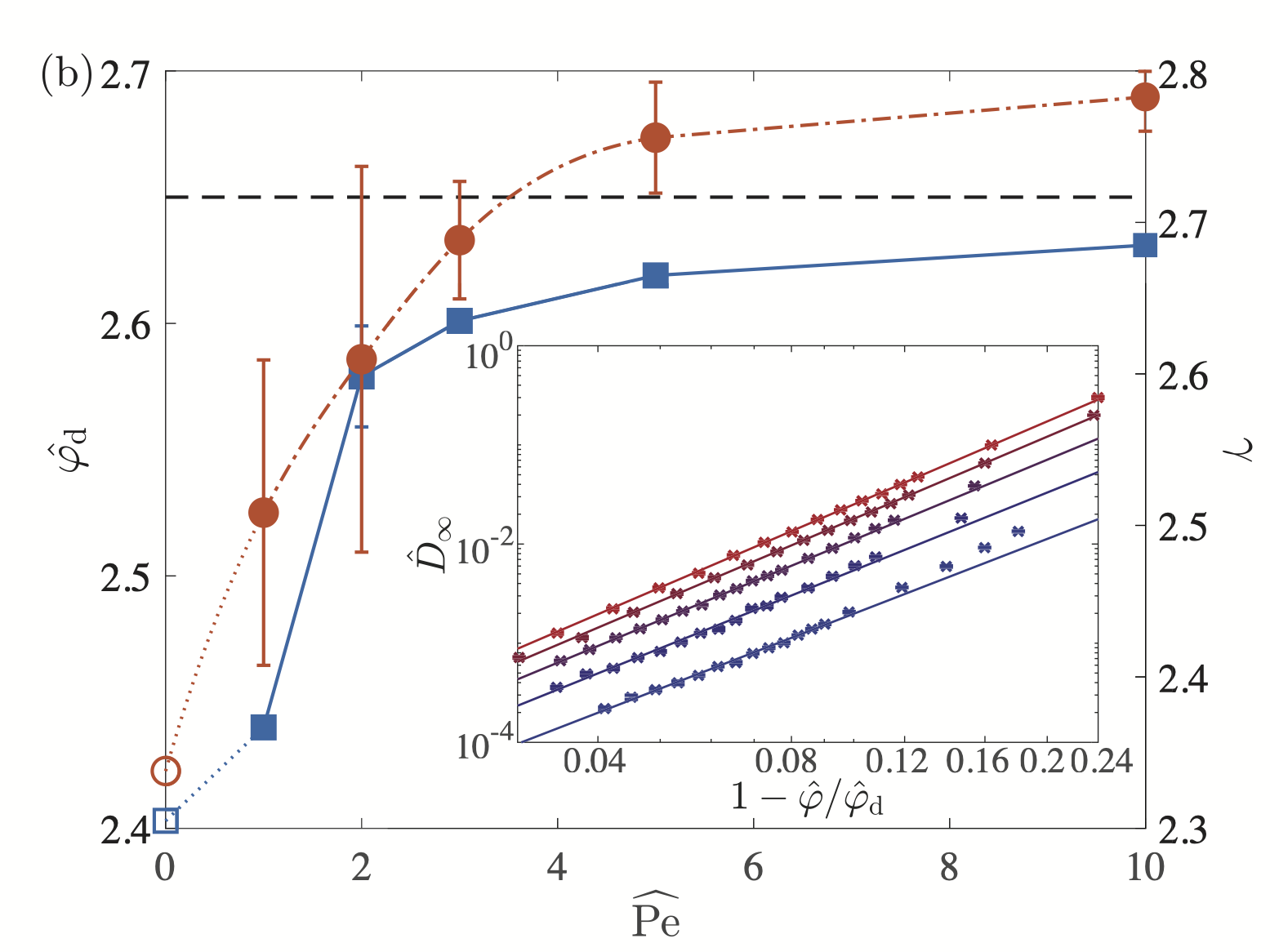}
\caption{(a) MSD for $\widehat{\mathrm{Pe}} = 10$ for $\widehat{\varphi} = 0.5, 1, 2, 2.5$, and $4$ (from top to bottom), and $d = 3, 4, 6$, and $12$ (fading as $d$ decreases) compared with the free-space result from Eq.~\eqref{eq:abp_msd_app} (black dashed line). (Insets) For $d = 12$,  (top) The ABP-RLG diffusivity at the same densities (lines from top to bottom) and (bottom) its ratio with the BP-RLG diffusivity. The long-time diffusivity of ABP-RLG is clearly enhanced. (b) The critical density $\widehat{\varphi}_\mathrm{d}$ for $d=12$ (squares) increases with $\widehat{\mathrm{Pe}}$ and appears to saturate around the percolation threshold $\widehat{\varphi}_\mathrm{p}(d=12)=2.64(9)$~\cite{biroli2021interplay} (black dashed line), and so does the scaling exponent $\gamma$ (circles). Results for $\widehat{\mathrm{Pe}}=0$ in the limit $d\rightarrow\infty$ -- $\widehat{\varphi}_\mathrm{d}=2.4034$ and $\gamma = 2.33786$ (empty symbols) -- are provided as reference~\cite{biroli2021interplay,kurchan2013exact}. Lines are guides for the eye. (Inset) Critical scaling of the long-time diffusivity $\widehat{D}_\infty$ for $d = 12$ with $\widehat{\mathrm{Pe}}=1,2,3,5,10$ (blue to red lines).}
\label{fig:msd}
\end{figure*}

For the BP-RLG, the MSD cleanly converges to the $d\rightarrow\infty$ DMFT results as $d$ increases~\cite{charbonneau2024dynamics,manacorda2020numerical}; for the ABP-RLG, DMFT predictions remain unavailable, but the dimensional trend is similar (see Fig.~\ref{fig:msd}(a)). As expected, the long-time diffusion is enhanced as the P\'{e}clet number grows (see Fig.~\ref{fig:msd}(a) inset). More surprisingly, this effect increases with density in the diffusive regime. In other words, when an abundance of obstacles geometrically hinders escapes, activity affects diffusion more by increasing the probability of exploring escape paths out of cages. Caging in active systems is therefore qualitatively different than in standard liquids (see also Fig.~\ref{fig:traj}). 

In order to quantify the resulting impact of this accelerated uncaging, we estimate the (avoided) dynamical transition $\widehat{\varphi}_\mathrm{d}$, at which the tracer would localize in an MCT-like description. The long-time diffusivity, $\widehat{D}_\infty = \lim\limits_{\hat{t}\rightarrow\infty}\frac{\widehat{\Delta}(\hat{t})}{\hat{t}}$,
is therefore taken to scale as 
\begin{equation}
\label{eq:scaling_diffusivity}
    \widehat{D}_\infty \sim \left\lvert {\widehat{\varphi}-\widehat{\varphi}_\mathrm{d}}\right\rvert^\gamma.
\end{equation}
Because for $d \leq 8$, the geometric percolation transition takes place below the mean-field localization, i.e., $\widehat{\varphi}_\mathrm{p}\leq \widehat{\varphi}_\mathrm{d}$~\cite{biroli2021interplay}, the two phenomena are then easily confounded. We here avoid this problem by considering the case $d = 12$, where $\widehat{\varphi}_\mathrm{p} = 2.64(9)$ is well above $\widehat{\varphi}_\mathrm{d}\approx 2.40$ for the BP-RLG~\cite{biroli2021interplay}. (The same holds for $d=10$~\cite{SI}.) In such high $d$, perturbative deviations (in $1/d$) from mean-field $d\rightarrow\infty$ physics are also rather small~\cite{charbonneau2024dynamics}. As expected from prior studies of active systems, the critical density grows with increasing P\'{e}clet number (Fig.~\ref{fig:msd}(b))~\cite{berthier2014nonequilibrium,liluashvili2017mode}. 
Interestingly, the effect eventually saturates at around 10\%. This gap is much larger than the $\sim 1\%$ predicted by MCT for $d=2$ systems with comparable activity~\cite{liluashvili2017mode}. The saturation further coincides with the percolation threshold, $\widehat{\varphi}_\mathrm{p}$, which provides a clear geometric upper bound to cage escapes. By preferentially sampling low-dimensional features of caging, the active tracer effectively sidesteps the entropic bottlenecks of the BP-RLG in this regime~\cite{biroli2021interplay}, and is therefore largely oblivious to these instantonlike hopping processes. 

\begin{figure*}[t]
\centering
\includegraphics[scale=0.6,trim={0.4cm 0cm 0.1cm 0.1cm},clip]{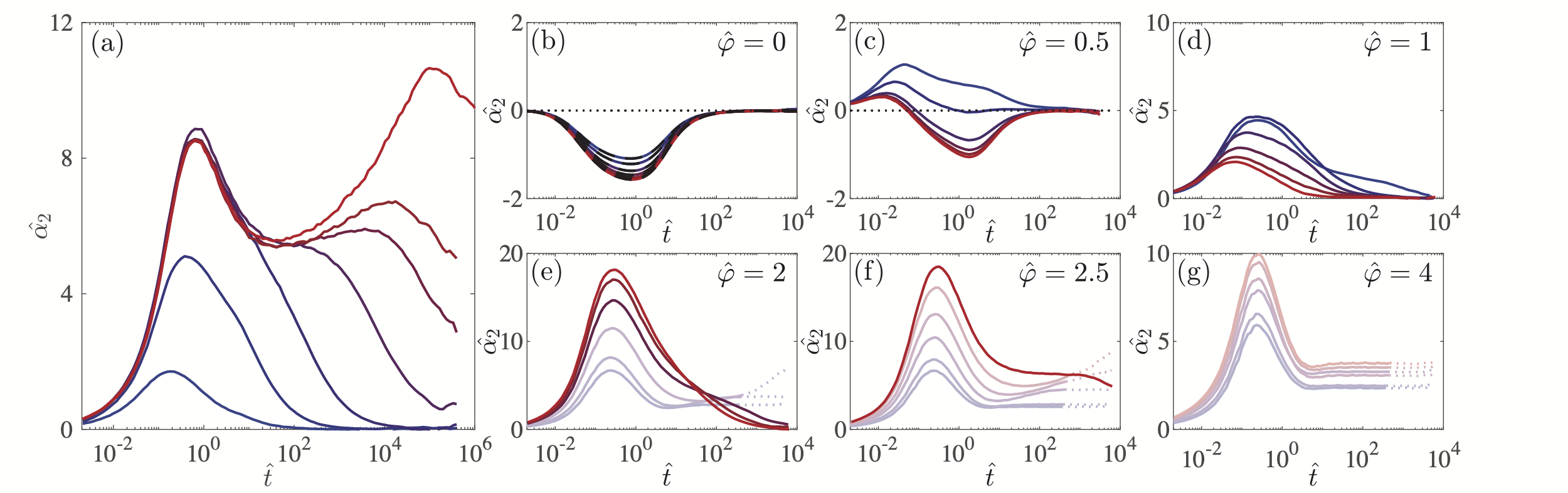}
\caption{Time evolution of $\widehat{\alpha}_2$ for the ABP-RLG with (a) $\widehat{\mathrm{Pe}} = 5$ and various obstacle densities $\hat{\varphi}=1, 1.5, 2, 2.3, 2.38, 2.4,$ and $2.42$ (blue to red lines), and (b-g) $\widehat{\mathrm{Pe}} = 10$ in $d = 3, 4, 6, 8, 10,$ and $12$ (blue to red lines) for given obstacle densities. For $\widehat{\varphi}=0$, analytical expressions (dashed lines) can be obtained in all $d$~\cite{SI}. Note that results for $\widehat{\varphi}\gtrsim\widehat{\varphi}_\mathrm{p}$, in which cases the traces does not diffuse at long times, are faded out to clarify the dimensional trend, and the ranges corresponding to percolation physics are dotted out.}
\label{fig:a2}
\end{figure*}

The behavior of the exponent $\gamma$ is also physically interesting. Although its precise numerical value is coupled to that of $\hat{\varphi}_d$ and therefore depends on the details of the fitting procedure~\cite{SI}, it nevertheless robustly increases with $\widehat{\mathrm{Pe}}$. This growing sensitivity of the diffusion to activity is markedly distinct from its invariance assumed in the analysis of spin-glass models~\cite{berthier2014nonequilibrium}, and opposite to recent MCT predictions~\cite{reichert2021mode}. Whether these calculations considered the relevant length scale or properly accounted for the impact of activity on the structure is, however, unclear, so we cannot exclude that the discrepancy could be resolved through a more careful analysis of the theory. In any event, the ABP-RLG results provide a stringent test of the impact of activity on sluggish dynamics.  

In order to probe the nature of active dynamics more thoroughly, we also consider its  fluctuations. A common measure of particle-scale dynamical heterogeneity in standard glass-forming liquids~\cite{rahman1964correlations,kob1997dynamical} is the non-Gaussian parameter (NGP), which in the RLG fully captures the effect~\cite{biroli2022local,folena2022equilibrium}, 
\begin{equation}
  \alpha_2(t)=  \frac{d}{d+2} \frac{[\langle r^4(t) \rangle]}{[\langle r^2(t) \rangle]^2} - 1.
\end{equation}
Here, $\langle\cdots\rangle$ denotes averaging over initial conditions in one realization of disorder and $[\cdots]$ denotes averaging over multiple realizations. The quantity is dimensionally rescaled as  $\widehat{\alpha}_2=d\cdot\alpha_2$~\cite{charbonneau2024dynamics}, thus capturing the small (perturbative) deviations from  mean-field $d\rightarrow\infty$ physics~\cite{charbonneau2024dynamics}. Deviations from $\widehat{\alpha}_2(t)=0$, which originate either from an enhanced displacement along a given direction or from directional correlations~\cite{biroli2022local}, are positive for a (van Hove\cite{hansen2006theory}) distribution of tracer displacements at fixed time with a large tail and negative otherwise. 
In passive systems, directional correlations result in the NGP markedly growing and peaking at \emph{times that grow upon approaching $\hat{\varphi}_d$}. In active systems, its behavior is markedly different.

Figure~\ref{fig:a2} shows that at both short and long times $\widehat{\alpha}_2$ vanishes for the ABP-RLG, as expected from the diffusive nature of transport in either regime. At intermediate times, the NGP of free particles turns \emph{negative}  due to the anti-correlation of displacement along different dimensions, as previously noted in Ref.~\cite{ten2011brownian}. Interestingly, just like the BP-RLG, the ABP-RLG exhibits a long-time peak which diverges as $\hat{\varphi}$ approaches $\hat{\varphi}_d$, as noted in Ref.~\cite{keta2022disordered}, but overlooked until then~\cite{ten2011brownian,zheng2013non,ding2017study}. Even more interestingly, a separate \emph{short-time} ($\hat{t}=\mathcal{O}(1)$) peak also develops and grows in that density regime. 
Keta et al.~\cite{keta2022disordered} have suggested that the former peak corresponds to the universal behavior of liquid glass formers while the latter reveals ``the nontrivial influence of many-body interactions in persistent liquids.'' The ABP-RLG results, however, clearly indicate that both effects are single particle in nature.

\begin{figure}[hbt]
\centering
\includegraphics[scale=0.47,trim={0.5cm 0.1cm 0.8cm 0.2cm},clip]{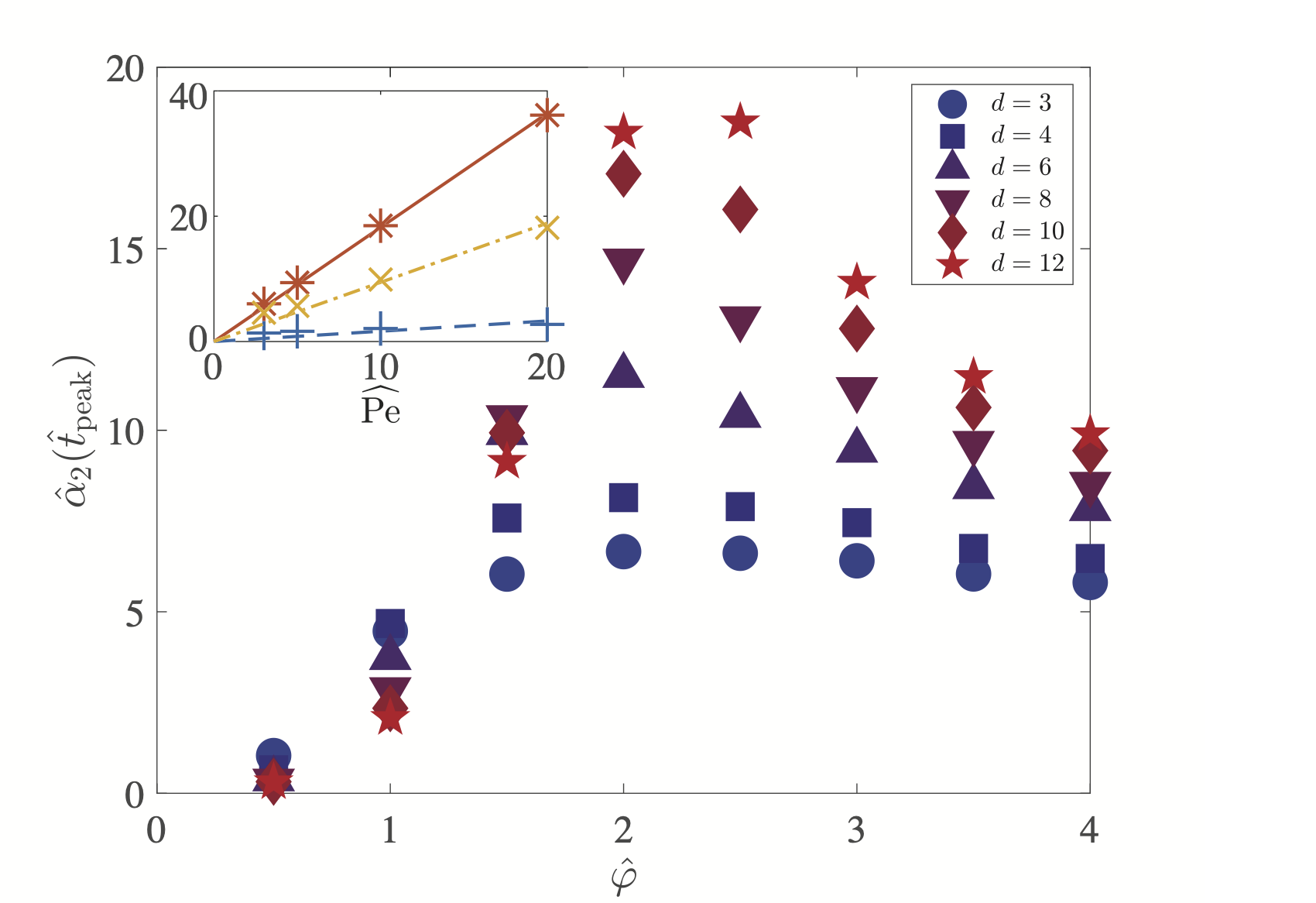}
\caption{Density evolution of the height of the first peak of $\hat{\alpha}_2$ for the ABP-RLG at $\widehat{\mathrm{Pe}}=10$ for various $d$ and obstacle densities. The peak height generally saturates as $d$ increases, but is roughly proportional to $d$ around $\hat{\varphi}_d$ similarly to the long-time peak of the BP-RLG~\cite{charbonneau2024dynamics}. (inset) The linear growth of the peak height with $\widehat{\mathrm{Pe}}$ for $d=12$ and $\hat{\varphi}=1$ (dashed line), $2.5$ (solid line) and $4$ (dot-dashed line).}
\label{fig:a2_peak1}
\end{figure}

The finite-time position of the first peak and its growth with dimension hint at a direct geometrical interpretation. Figure~\ref{fig:a2_peak1} reveals that the peak height first increases with $\hat{\varphi}$, saturates around $\hat{\varphi}_d$, and then decreases.  It also shows that $\widehat{\mathrm{Pe}}$ linearly enhances the effect (inset)~\cite{SI}. Because $\hat{\alpha}_2$  encodes fluctuations in cage sizes, which are maximal around $\hat{\varphi}_d$~\cite{biroli2022local,charbonneau2024dynamics}, the explanation for the peak position naturally follows. It further highlights the critical origin of the phenomenon, with an actual divergence expected in the limit $d\rightarrow\infty$. The enhancement of the effect with $\widehat{\mathrm{Pe}}$ is consistent with an active tracer exploring ever lower dimensional features -- down to edges and cavities formed by obstacles -- of its surroundings upon approaching caging. The robustness and magnitude of this effect suggest its importance in capturing the dynamics of active glass formers. A proper theory of these systems would then be expected to capture this distinctive fluctuation-based signature. 

\paragraph{Conclusion --}
The ABP-RLG provides key insights into the physics of dense active systems. First, its mean-field--like dynamical transition shifts to higher densities with activity and saturates at the percolation transition, thus identifying activity as a mechanism that sidesteps the instantonlike (or activated) hopping processes of standard glass formers.  Second, an active tracer tends to  explore lower-dimensional caging features,  resulting in a clear and distinct short-time peak of the NGP around  $\widehat{\varphi}_\mathrm{d}$. This finite-dimensional echo of mean-field criticality is well within the reach of simulation and experiments in these systems and is therefore a distinctive feature with which to contrast standard and active glass formers. These findings could further be used to design systems in which the efficacy of active particle caging can be tuned, especially for $d>2$. Finally, these findings once again highlight the importance of solving the DMFT, as well as its small and large fluctuation corrections, as the necessary path toward achieving a first-principle description of glasses, both active and not.

\begin{acknowledgments}
\paragraph{Acknowledgments--}
We thank Ludovic Berthier, Giampaolo Folena, Yi Hu, Grzegorz Szamel, and Yikang Zhang for many stimulating exchanges. This work was supported by a grant from the Simons Foundation (Grant No.~454937). D. K. thanks the European Union Horizon 2020 research and innovation program under Grant Agreement No. 857470 and from the European Regional Development Fund via the Foundation for Polish Science International Research Agenda PLUS program GrantNo. MAB PLUS/2018/8, and the Ministry of Science and Higher Education’s initiative “Support for the Activities of Centers of Excellence Established in Poland under the Horizon 2020 Program” (agreement no. MEiN/2023/DIR/3795
\paragraph{Data availability --} Data relevant to this work have been archived and can be accessed at the Duke Research Data Repository \cite{dataduke}.
\end{acknowledgments}


\bibliographystyle{apsrev4-1}
\bibliography{reference}

\end{document}


\beginsupplement
\title{\Large Supplementary Information for\\ ``Not-so-glass-like Caging and Fluctuations of an Active Matter Model''}

\author{\small Mingyuan Zheng,\textit{$^{a,}$}$^{\ast}$ Dmytro Khomenko,\textit{$^{b}$} and Patrick Charbonneau\textit{$^{a,c}$}}

\date{{\footnotesize \em $^a $Department of Chemistry, Duke University, Durham, North Carolina 27708, USA\\
$^b$Dipartimento di Fisica, Sapienza Universit\`a di Roma, Piazzale Aldo Moro 5, 00185 Rome, Italy\\
$^c$Department of Physics, Duke University, Durham, North Carolina 27708, USA\\
$^*$ Email: mingyuan.zheng@duke.edu\\
\today}}
\maketitle

\section{Simulation Details}
The event-driven Brownian dynamics (EDBD) algorithm is here adapted to the ABP-RLG system~\cite{foffi2005scaling,scala2007event,charbonneau2024dynamics}. Special attention must then be given to wedge escapes.

\subsection{EDBD Algorithm in ABP-RLG}
Simple Brownian dynamics simulations are run with an integration time step $\Delta t_n = 2^n \Delta t_0$, where $\Delta t_0$ is the smallest time interval, and for each $n$, motion within $\Delta t_n$ is sampled for $2^{10}$ times; at the end of each $\Delta t_n$, the translational velocity is reset as a multi-variate Gaussian random variable. Because active motion calls for a smaller time interval to approximate the linear component of motion, we set $\Delta t_a = \Delta t_n/m$ (where $m$ is a positive integer depending on dimensionality, P\'{e}clet number and density), and the direction of the rotational velocity is updated with Gaussian random fluctuation at the end of each $\Delta t_a$. 

To minimize the necessary system size, checkerboard periodic boxes are considered  for $d = 3, 4, \cdots, 12$ as in Ref.~\cite{charbonneau2021high}. The system is initialized by placing a tracer at the origin with a Gaussian random velocity, and $N = 100,000$ obstacles that do not overlap with the tracer and are otherwise randomly distributed so as to fix $\hat{\varphi}$. 
The system is sampled after an initialization time $t = 100$, so as to attain stationary states, and the results are averaged over $2000$ realizations.

\subsection{Escape velocity}
\begin{figure}[b]
\centering
\includegraphics[scale=0.45,trim={0.8cm 0 0.1cm 0.2cm},clip]{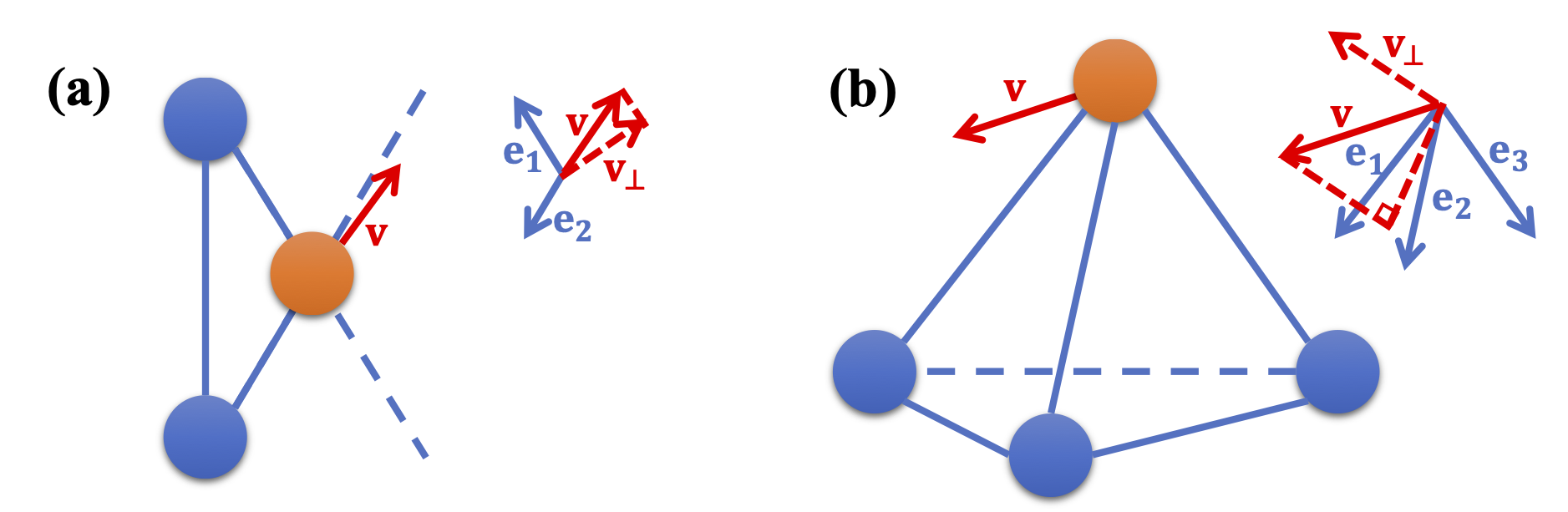}
\caption{Schematics of determining escape velocity $\mathbf{v_\bot}$ of a tracer in contact with $n=d$ obstacles in (a) $d=2$ and (b) $d=3$. In (a), $\mathbf{v} = x_1\mathbf{e_1}+x_2\mathbf{e_2}$ has $x_1,x_2 < 0$, while the tracer \textit{squeezes} one of the obstacles and thus $\mathbf{v_\bot}$ is perpendicular to $\mathbf{e_1}$. In (b), $\mathbf{v} = \sum\limits_{i=1}^d x_i\mathbf{e_i}$ has $x_1,x_2 > 0$ and $x_3 < 0$, and 
$\mathbf{v_\bot}$ is perpendicular to the surface of the simplex formed by $\mathbf{e_1}$ and $\mathbf{e_2}$.}
\label{fig:si_v}
\end{figure}

The escape velocity of an active tracer blocked by obstacles is determined as follows. Given a tracer in contact with $n$ obstacles (in $d$-dimensional space, with probability one a tracer is in contact with at most $d$ obstacles, thus forming a simplex), and $\mathcal{K} = \{\mathbf{e_i}, \,i = 1\cdots n\}$  non-orthogonal unit vectors denoting the directions from the tracer to obstacles in contact. When $n = d$, $\{\mathbf{e_i}\}$ forms a complete set of basis vectors, and the original velocity can be expressed as $\mathbf{v} = \sum\limits_{i=1}^d x_i\mathbf{e_i}$. The escape condition is that there exists at least one coefficient $x_i < 0$, which means the direction of the original velocity is out of the trap angle. When $n < d$, the possibility of being trapped is non-trivial. Determining the escape velocity, $\mathbf{v_\bot}$, is then essential but not as simple, because $\mathbf{v_\bot}$ can be perpendicular to lines, surfaces, $\cdots$ $(d-1)$-order geometrical features of the simplex -- if $n<d$, at most $n$-order features. Contrary to (naive) intuition, the direction of $\mathbf{v_\bot}$ cannot be directly obtained from the coefficients $\{x_i\}$ of the aforementioned linear combination, but is related instead to the angle between the original velocity $\mathbf{v}$ and the features of the (incomplete) simplex, as illustrated by the example of $n=d=2$ shown in Fig.~\ref{fig:si_v}(a). 

In order to determine the escape velocity $\mathbf{v_\bot}$ for general cases $n\leq d$, we define
\begin{equation}
    \mathbf{v} = \mathbf{v_\bot}^{(j)} + \sum\limits_{j} x_j\mathbf{e_j},
\end{equation}
where all possible nonempty subsets $\{\mathbf{e_j}\}\subseteq \mathcal{K}$ are considered. To obtain $\mathbf{v_\bot}^{(j)}$ such that perpendicular to all basis vectors in $\{\mathbf{e_j}\}$ becomes a least-squared-approximation (LSA) problem solving $Ax = b$. The resulting $\mathbf{v_\bot}$ from the possible solutions of $\mathbf{v_\bot}^{(j)}$ fulfills the following conditions: (1) the corresponding coefficients $\{x_j\}$ are all positive, which means that this subset $\{\mathbf{e_j}\}$ is a real obstacle feature; (2) $\mathbf{v_\bot}$ does not \textit{squeeze} any of the obstacles in contact, which means that $\mathbf{v_\bot}\cdot\mathbf{e_i}\leq 0$ for 
$\mathbf{e_i}\in\mathcal{K}$); and (3) $\mathbf{v_\bot}$ has the minimal norm among all possible solutions. A schematic of this construction for $d=3$ is shown in Fig.~\ref{fig:si_v}(b). 

\section{Choosing $D_r/D_t$ in the ABP-RLG}
We here clarify the determination of the dimension-dependent rescaling parameters for time, MSD and NGP, which is found to also depend on $D_r/D_t$ in the ABP system. We particularly explore the effects of $D_r/D_t$ in both free space and heterogeneous environments, and the choice of $D_r/D_t$ does not change the underlying physics. The understanding obtained from the ABP model with $D_r/D_t=d$ is universal, and the results are comparable to those from Ref.~\cite{keta2022disordered}, where $D_t = 0$.

\begin{figure}[h!]
\centering
\includegraphics[scale=0.42,trim={0.2cm 0 0.1cm 0.2cm},clip]{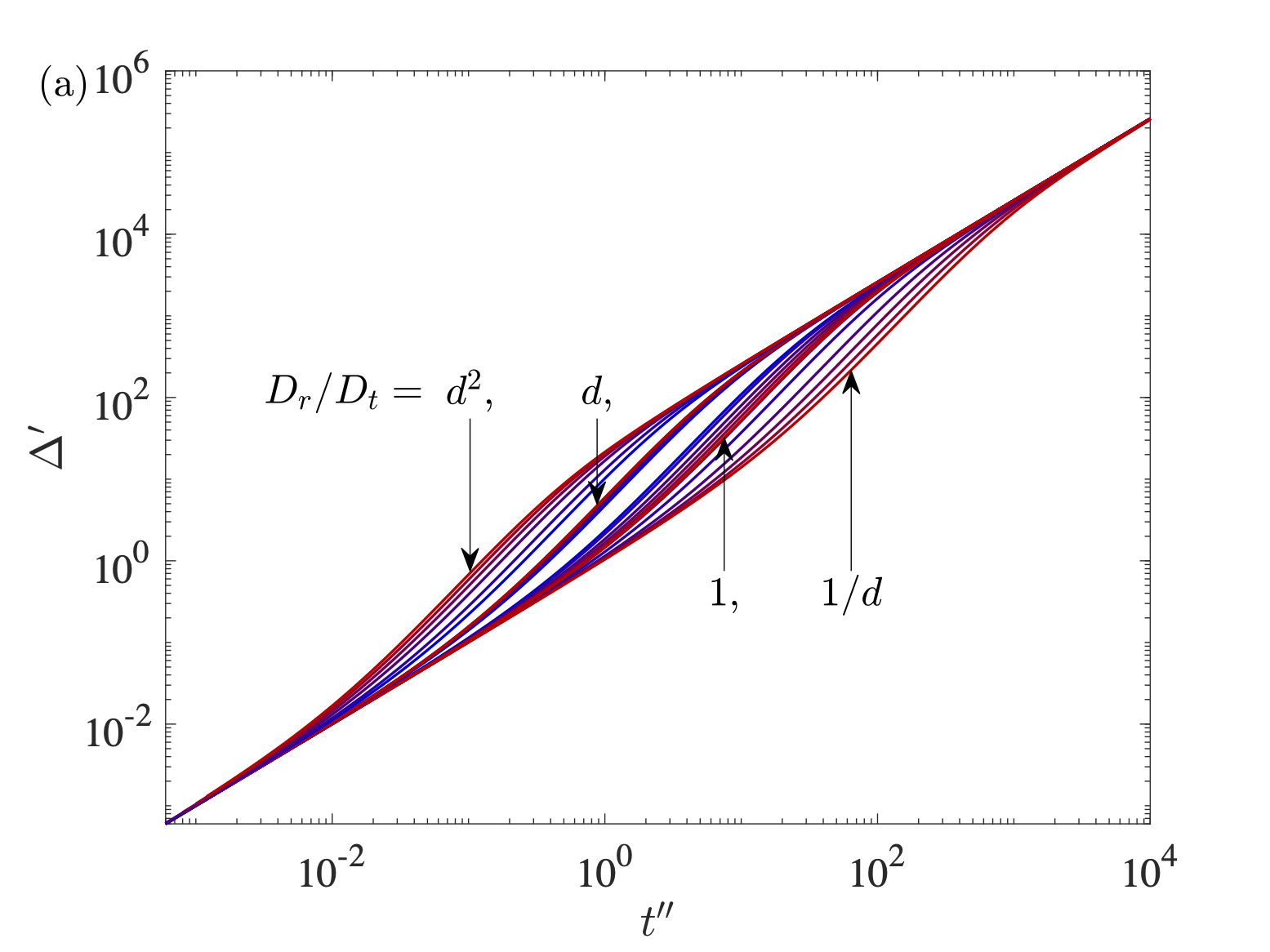}
\includegraphics[scale=0.42,trim={0.2cm 0 0.1cm 0.2cm},clip]{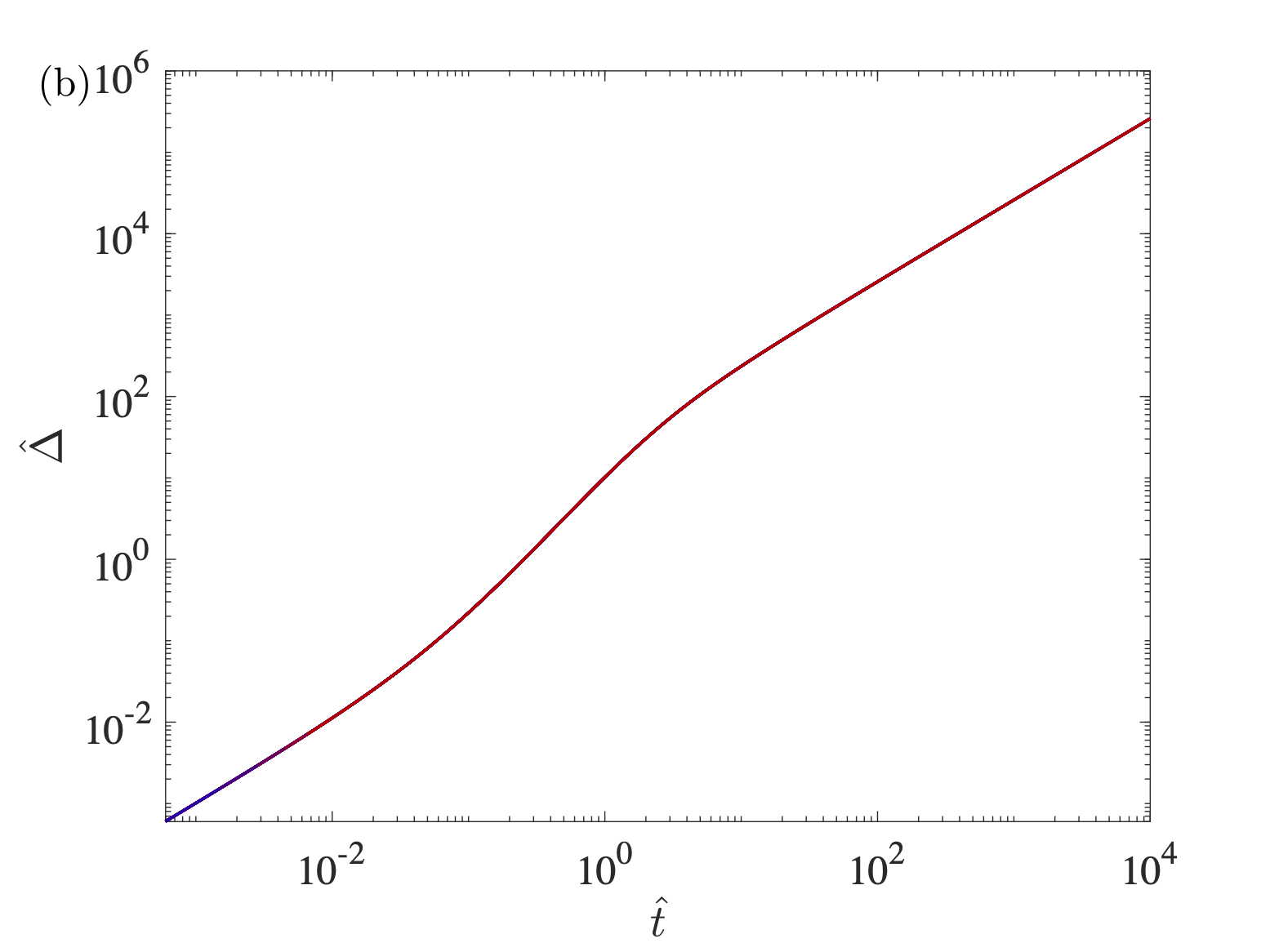}
\includegraphics[scale=0.42,trim={0.2cm 0 0.1cm 0.2cm},clip]{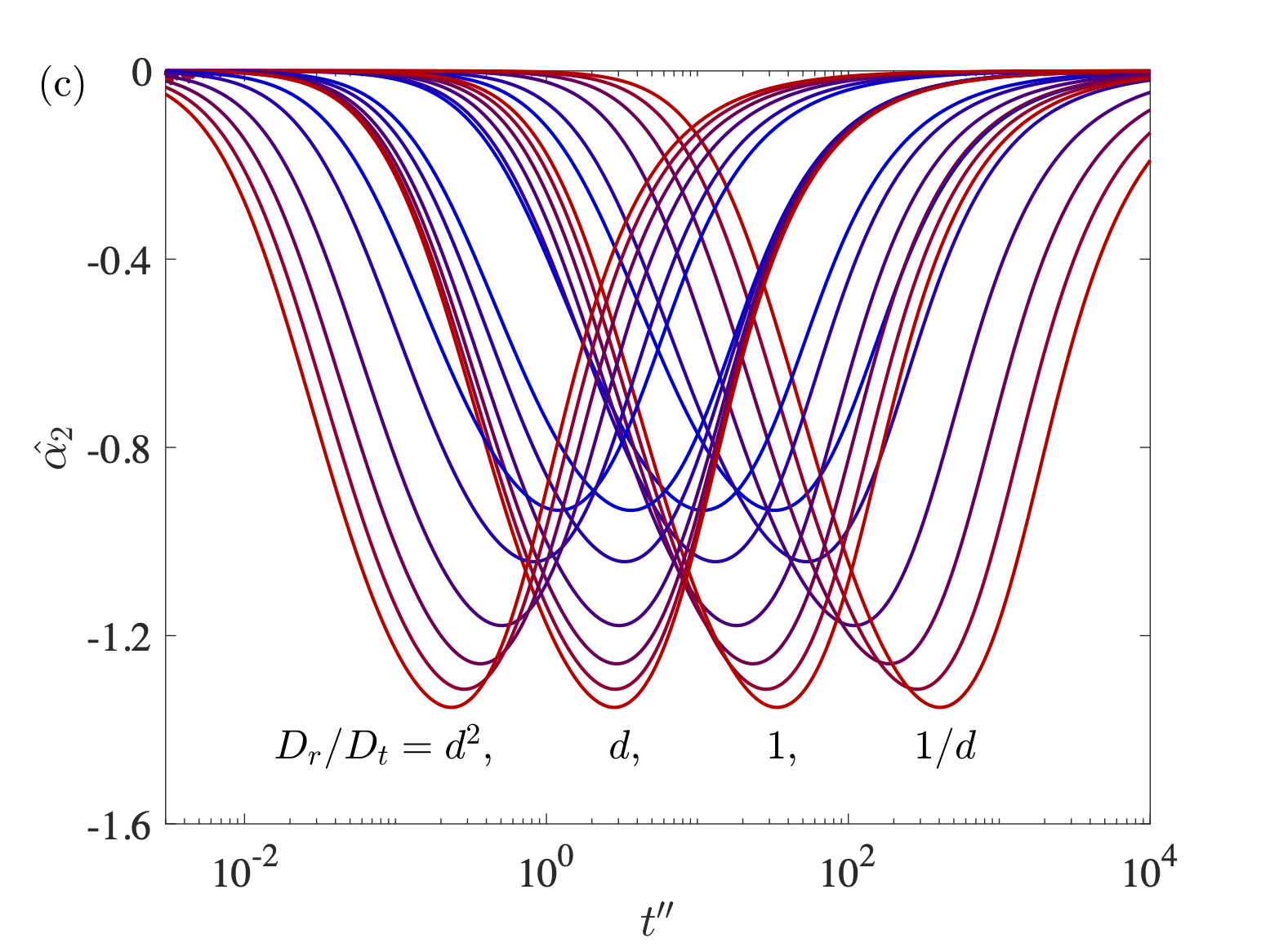}
\includegraphics[scale=0.42,trim={0.2cm 0 0.1cm 0.2cm},clip]{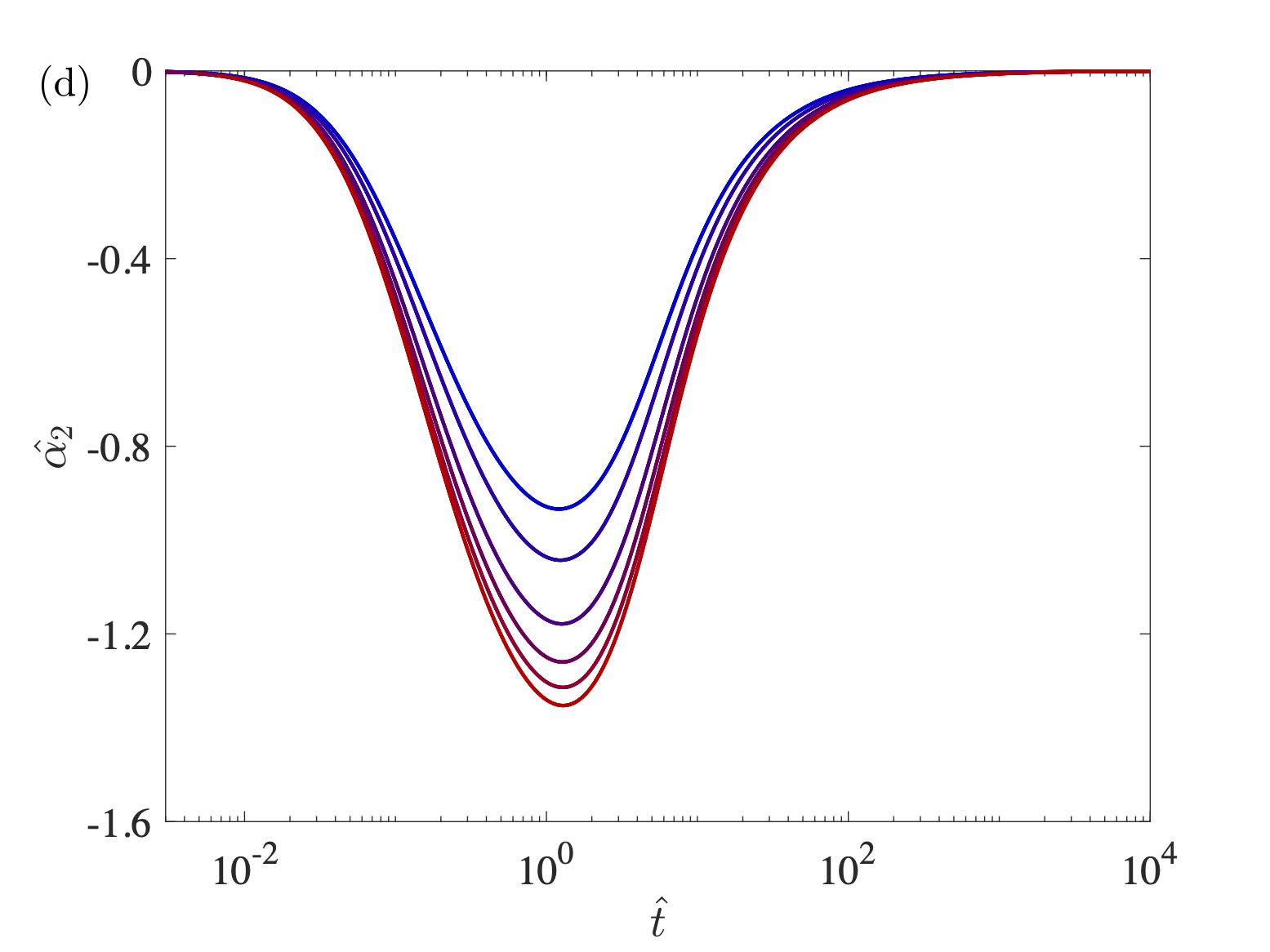}
\caption{Theoretical (a-b) MSD and (c-d) $\hat{\alpha}_2(\hat{t})$ of an ABP in free space for $d = 3, 4, 6, 8, 10, 12$ (blue to red lines), at $\widehat{\mathrm{Pe}}=5$, and $D_r/D_t = d^2, d, 1, 1/d$. Time and MSD both are rescaled by $d$ in (a) and (c), and by $wd$ in (b) and (d) ($\hat{\alpha}_2$ is as defined). In (b), the MSDs of different dimension and different $D_r/D_t$ overlap. In (d), the $\hat{\alpha}_2$ well collapses with $d$, and that of the same $d$ is invariant with $D_r/D_t$.}
\label{fig:si_freespace}
\end{figure}

\subsection{Results of $\hat{\varphi}=0$}
Starting from the MSD of an ABP in free space (Eq.~(3) in the main text), 
we first consider $\Delta' = \Delta d$ and $t' = td$ are the dimensionally rescaled quantities as those of BP systems~\cite{charbonneau2024dynamics}, and $t''  = 2dD_t t'$ sets the time units,
\begin{equation}
\label{eq:msd}
    \Delta' = t'' + \frac{1}{w}\widehat{\mathrm{Pe}}^2(wt''-1+e^{-wt''})
\end{equation}
where $\widehat{\mathrm{Pe}}$ is as defined in the main text and $w = \frac{(d-1)D_r}{2d^2D_t}$ is about the dimension $d$ and the ratio $D_r/D_t$.

The rescaled P\'{e}clet number then collapses the short- and long-time diffusion regimes in all $d$,
\begin{equation}
\label{eq:abp_msd_app}
    \Delta' = \left\{ \begin{array}{ll}
    t'' + \frac{1}{2}w\widehat{\mathrm{Pe}}^2t''^2  \qquad\mbox{for $t''\ll 1$};\\
    (1+\widehat{\mathrm{Pe}}^2)t'' \qquad\quad\ \mbox{for $t''\gg 1$}.
    \end{array} \right.
\end{equation}
Here, the long-time diffusive regime gives the effective diffusive coefficient of the active particle $D_{\mathrm{eff}}/D_t = 1+\widehat{\mathrm{Pe}}^2$, where $\widehat{\mathrm{Pe}}$ measures the activity strength. However, $D_r$ and $v_0$ are both independent variables. In the main text, we set $D_r = dD_t$ and then only change $v_0$. (In Eq.~\eqref{eq:msd}, this corresponds to setting $w$ and changing $\widehat{\mathrm{Pe}}$.) To ensure that this choice does not qualitatively affect our conclusions, the role of $D_r$ should be more specifically considered.

We first fix $\widehat{\mathrm{Pe}}$ and change $w$ (and hence $D_r/D_t$). Fig.~\ref{fig:si_freespace} (a) shows the resulting MSDs for different $D_r/D_t$ ratios. The results are then inconsistent at intermediate times. More significantly, the corresponding $\hat{\alpha}_2$ (see Fig.~\ref{fig:si_freespace} (c) for expressions in Sec.~\ref{sec:si_a2}) have the same peak heights but shifted positions. Following Eq.~\eqref{eq:msd}, it is clear that time should be rescaled with a parameter related to $D_r/D_t$, probably by $w$. 

Setting $\hat{t} = wt''$ (and $\hat{\Delta}=w\Delta'$, accordingly), we obtain Eq.~(6) in the main text. The resulting MSDs are then invariant with $d$ and $D_r/D_t$; the $\hat{\alpha}_2$ for a given $d$ but different $D_r/D_t$ now overlap, and the peak positions for different $d$ no longer drift.

\subsection{Results of $\hat{\varphi}\not=0$}
\begin{figure}[t!]
\centering
\includegraphics[scale=0.47,trim={0.2cm 0 1.2cm 0.2cm},clip]{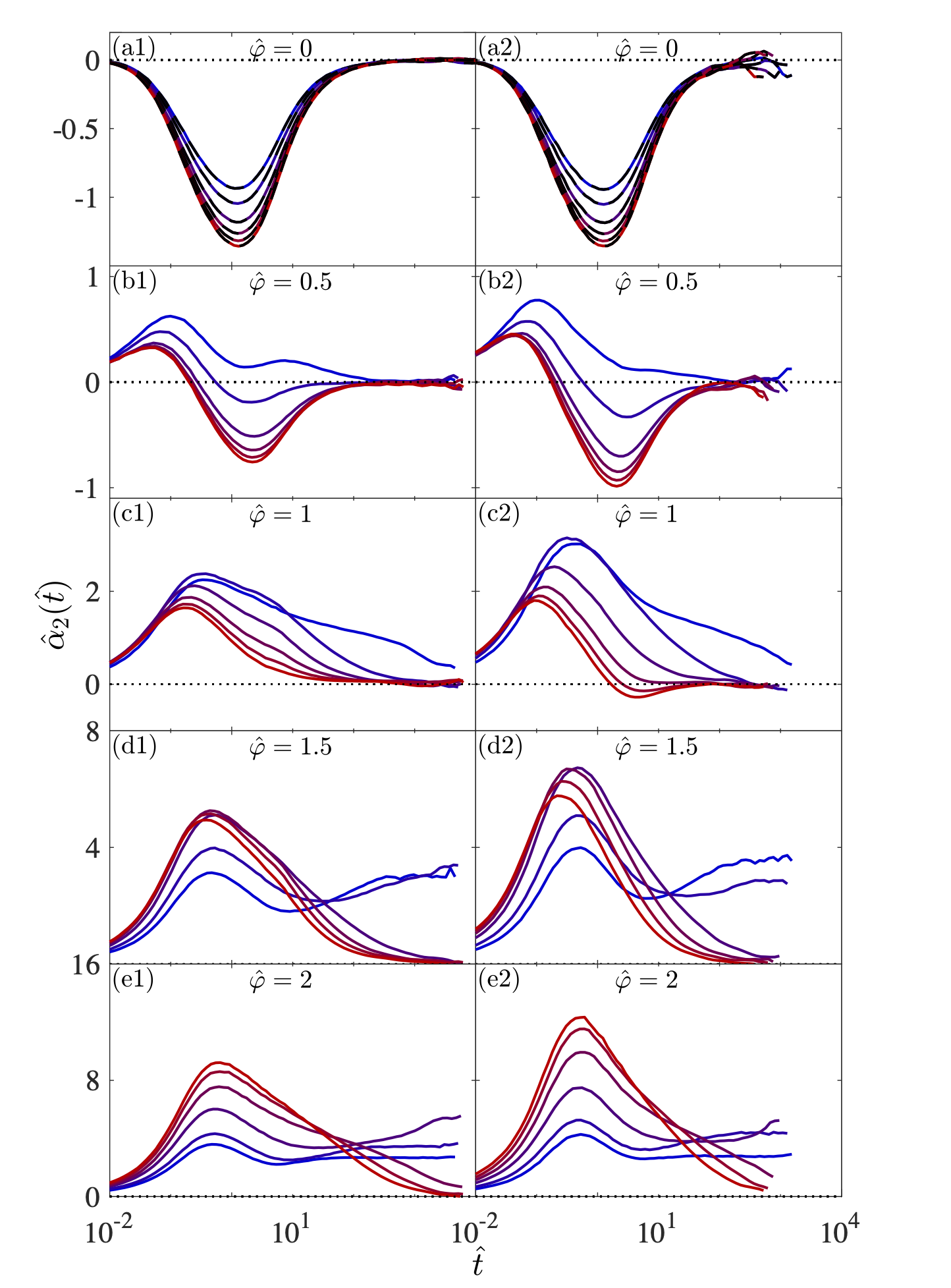}
\includegraphics[scale=0.47,trim={0.2cm 0 0.1cm 0.2cm},clip]{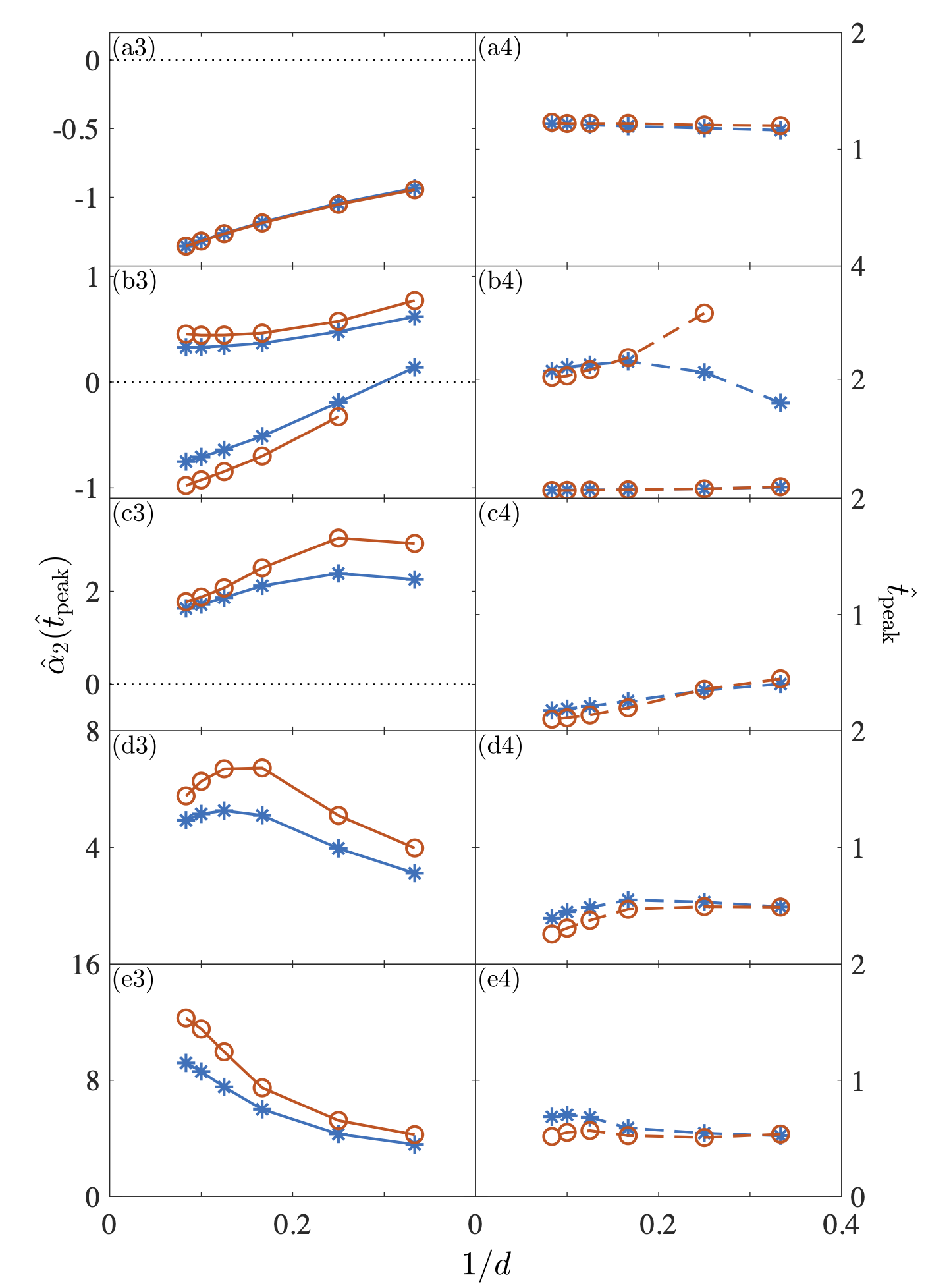}
\caption{(left) Time evolution of $\widehat{\alpha}_2(\hat{t})$ of the ABP-RLG for $D_r/D_t = d$ (a1)-(e1) and $1$ (a2)-(e2), and various $\widehat{\varphi}$ in $d = 3, 4, 6, 8, 10, 12$ (blue to red curves). For $\widehat{\varphi}=0$, simulation results are consistent with analytical expressions (dashed lines). Note that $\hat{\varphi}_p (d=6) = 2.022$, $\hat{\varphi}_p (d=4) = 1.562$, $\hat{\varphi}_p (d=3) = 1.17$~\cite{biroli2021interplay}, and therefore percolation physics affects the results for $\hat{\varphi}=2$ in $d\leq 6$, for $\hat{\varphi}=1.5$ in $d\leq 4$ as well as for $\hat{\varphi}=1$ in $d= 3$, especially at long times. (right) Dimensional evolution of the peak height (a3)-(e3) and peak position (a4)-(e4), for $D_r/D_t = d$ (blue stars) or $1$ (orange circles). For $\hat{\varphi}=0.5$, the positive and negative peaks are respectively estimated. For $\hat{\varphi}\leq 1.5$, the peaks collapse with $d$. For $\hat{\varphi} = 2$, the peak  heights of $d \leq 6$ already saturate (i.e. the peak height no longer increases with densities), and those of higher dimension show a trend of collapse.}
\label{fig:si_a2_phi}
\end{figure}

Fig.~\ref{fig:si_a2_phi} provides the simulation results of $\hat{\alpha}_2$ for $D_r/D_t$ equal to $d$ or $1$, with appropriate rescaling quantities. It indicates that in the environments of obstacles, the peak position is always invariant with $D_r/D_t$ regarding the fitting error, while the peak height increases as $D_r/D_t$ decreases. As the persistence time increases, the effect plays a more important role in the presence of obstacles than in free space. However, it does not change the underlying physics, so a single choice of $D_r/D_t$ suffices.

\section{NGP in free space}
\label{sec:si_a2}
The definitions of $\alpha_2(t)= \frac{d}{d+2} \frac{[\langle r^4(t) \rangle]}{[\langle r^2(t) \rangle]^2} - 1$ and $\langle r^2(t) \rangle^2$ for a free space ABP are given in the main text. Here, we consider $\angles{\mathbf{r}^4(t)}$ for an ABP in arbitrary $d$, building on the results for a pure translational Brownian particle and pure rotational Brownian particle~\cite{huang2015non,zheng2013non,sevilla2021generalized}. The resulting expression $\angles{\mathbf{r}^4(\hat{t})} = \angles{\mathbf{r}^4(\hat{t})}_1 + \angles{\mathbf{r}^4(\hat{t})}_2 + \angles{\mathbf{r}^4(\hat{t})}_3$, is naturally broken in three parts
\begin{align}
    w^2d^2\angles{\mathbf{r}^4(\hat{t})}_1 &= \frac{d+2}{d} \hat{t}^2,\\
    w^2d^2\angles{\mathbf{r}^4(\hat{t})}_2 &= \frac{2(d+2)}{d} \widehat{\mathrm{Pe}}^2 \hat{t}\left(\hat{t} - 1 + e^{-\hat{t}} \right),\\
    w^2d^2\angles{\mathbf{r}^4(\hat{t})}_3 &= \widehat{\mathrm{Pe}}^4 \left\{\frac{(d-1)^5}{d^3(d+1)^2}\,e^{-\frac{2d}{d-1}\hat{t}} - \frac{2}{(d+1)^2}\,e^{-\hat{t}}\left[-(d-7)(d+1)\hat{t} + (d+5)^2\right]\right. \\
    &+ \left.\frac{d+2}{d}\hat{t}^2 - \frac{2d^2+12d-2}{d^2}\hat{t} + \frac{d^3+23d^2-7d+1}{d^3}
    \right\},
\end{align}
where the first corresponds to translational motion, the third to rotational motion, and the second to their coupling. The resulting $\hat{\alpha}_2$ for an ABP in free space matches simulation results in Fig.~3.

\begin{figure}[t!]
\centering
\includegraphics[scale=0.45,trim={0.4cm 0 0.1cm 0.2cm},clip]{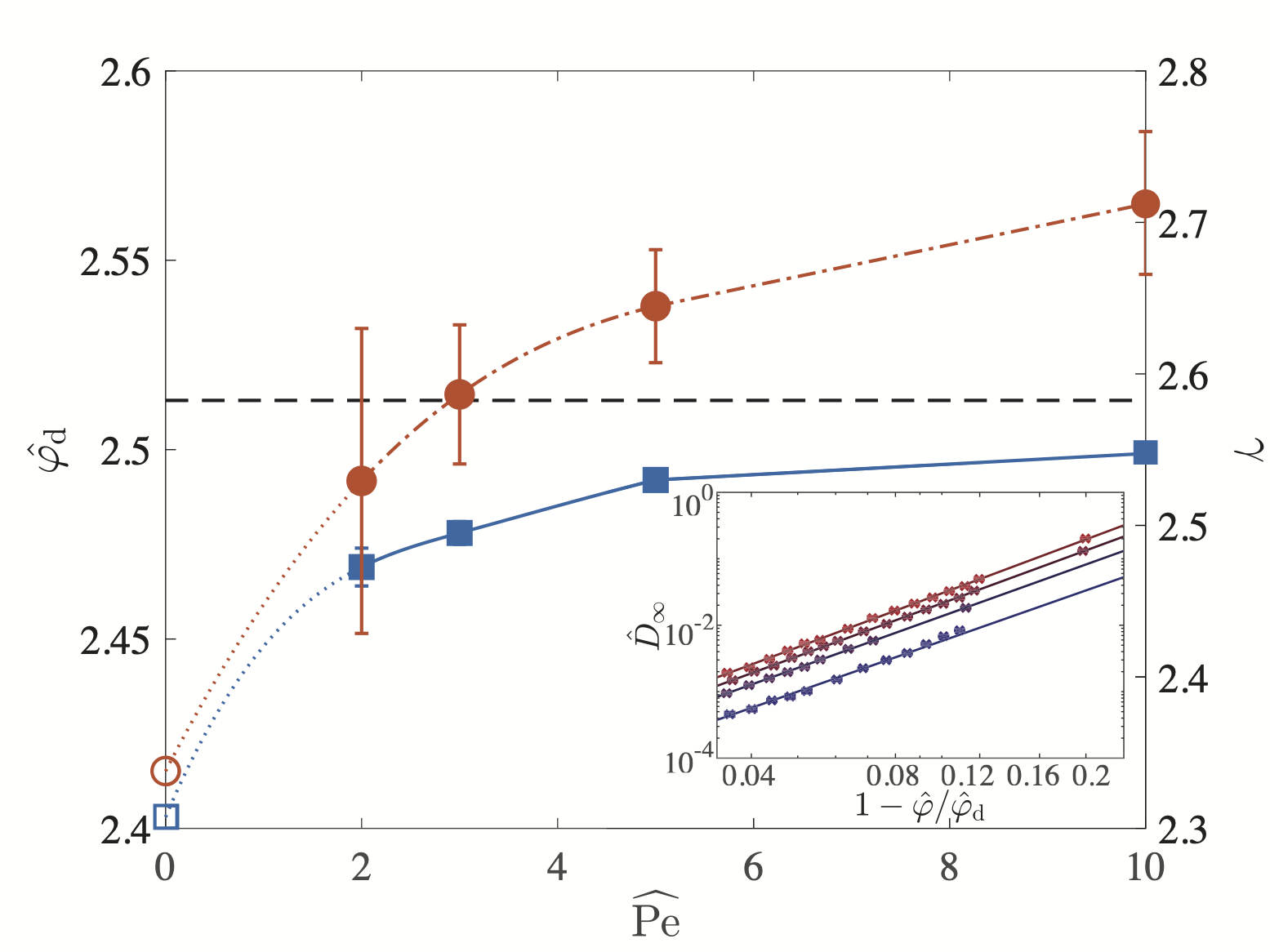}
\caption{The critical density $\widehat{\varphi}_\mathrm{d}$ (squares) increases with $\widehat{\mathrm{Pe}}$ and saturates around $\widehat{\varphi}_\mathrm{p}(d=10)=2.51(3)$~\cite{biroli2021interplay} (black dashed line), and so does the scaling exponent $\gamma$ (circles). Results for $\widehat{\mathrm{Pe}}=0$ in the limit $d\rightarrow\infty$ -- $\widehat{\varphi}_\mathrm{d}=2.4034$ and $\gamma = 2.33786$ (empty symbols) -- are provided as reference~\cite{biroli2021interplay,kurchan2013exact}. (Inset) Critical scaling of the diffusivity $\widehat{D}_\infty$ for $d = 10$ with $\widehat{\mathrm{Pe}}=2,3,5,10$ (blue to red lines).}
\label{fig:si_diff_d10}
\end{figure}

\section{Determination of the scaling exponent $\gamma$}
The long-time diffusivity $\hat{D}_\infty$ is obtained by fitting with the empirical form~\cite{charbonneau2022dimensional}
\begin{equation}
    \hat{\Delta}/\hat{t} = \hat{D}_\infty + a_0 t^{-a_1},
\end{equation}
where $a_0$ and $a_1$ are positive, but are otherwise not reported because their value plays no role in the subsequent analysis. 

\section{Results for $d=10$}
In order to ensure the trends of $\hat{\varphi}_d$ and $\gamma$ with $\widehat{\mathrm{Pe}}$ are not specific for $d=12$,  we here report a similar analysis for $d=10$, where also $\hat{\varphi}_p > \hat{\varphi}_d$ at $\widehat{\mathrm{Pe}}=0$. As can be seen in Fig.~\ref{fig:si_diff_d10}, the trends are essentially the same as in Fig.~2(b). The results for $d=12$ are therefore fully representative of the impact of activity on glass formation.

\section{Short-time peak height in $\alpha_2$ with $\widehat{\mathrm{Pe}}$}
\begin{figure}[t!]
\centering
\includegraphics[scale=0.45,trim={0.8cm 0 0.1cm 0.2cm},clip]{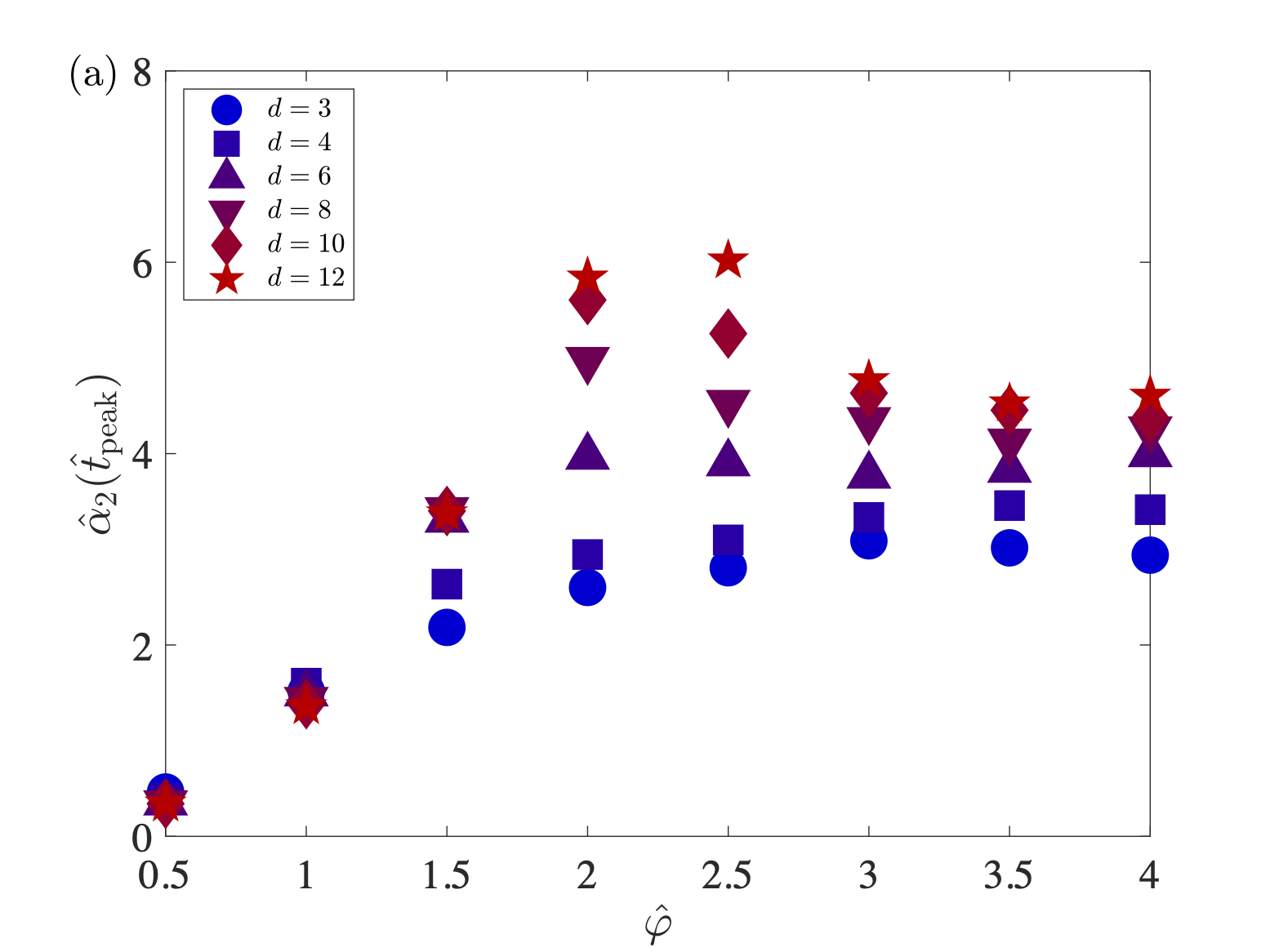}
\includegraphics[scale=0.45,trim={0.8cm 0 0.1cm 0.2cm},clip]{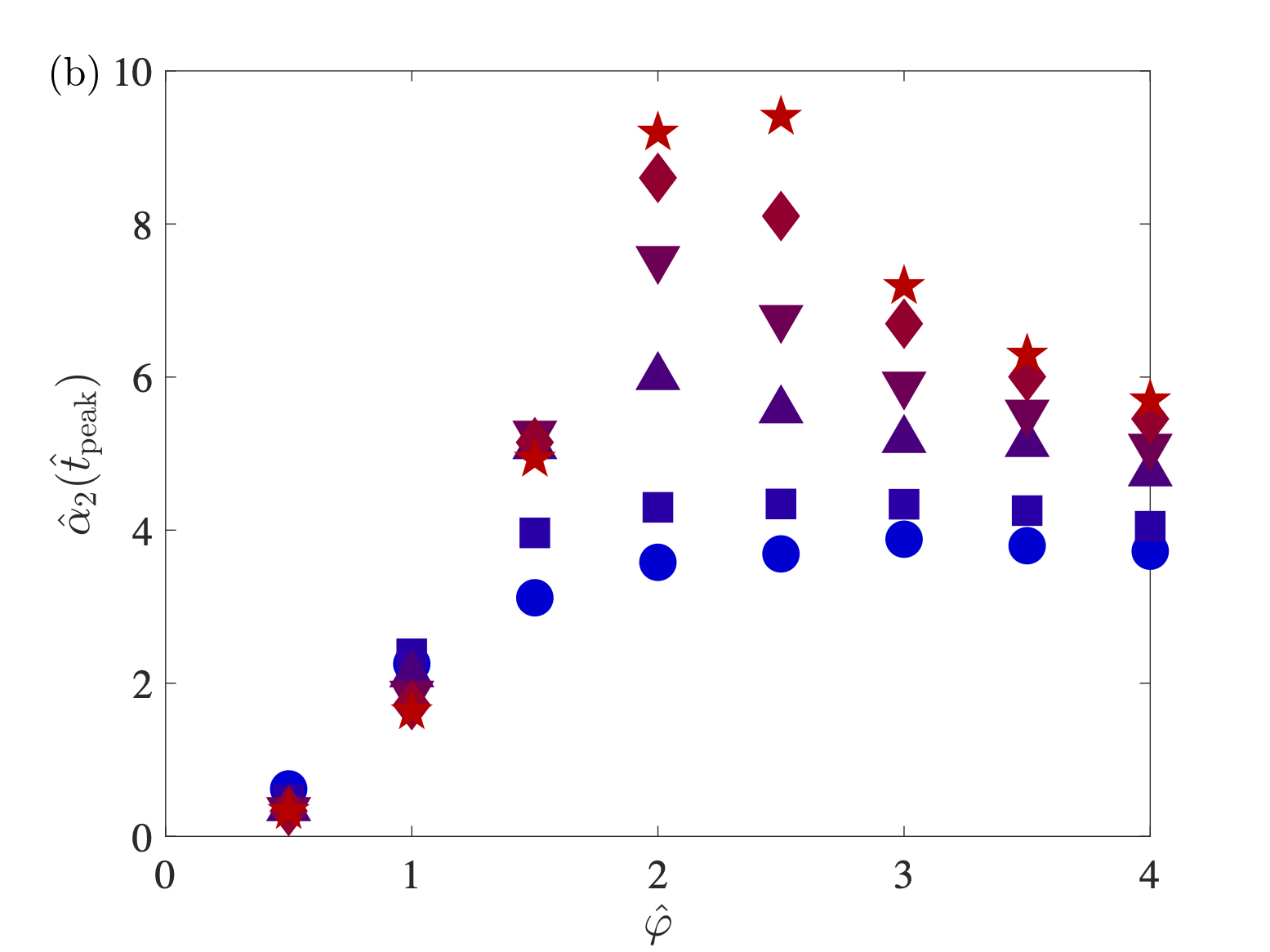}
\includegraphics[scale=0.45,trim={0.8cm 0 0.1cm 0.2cm},clip]{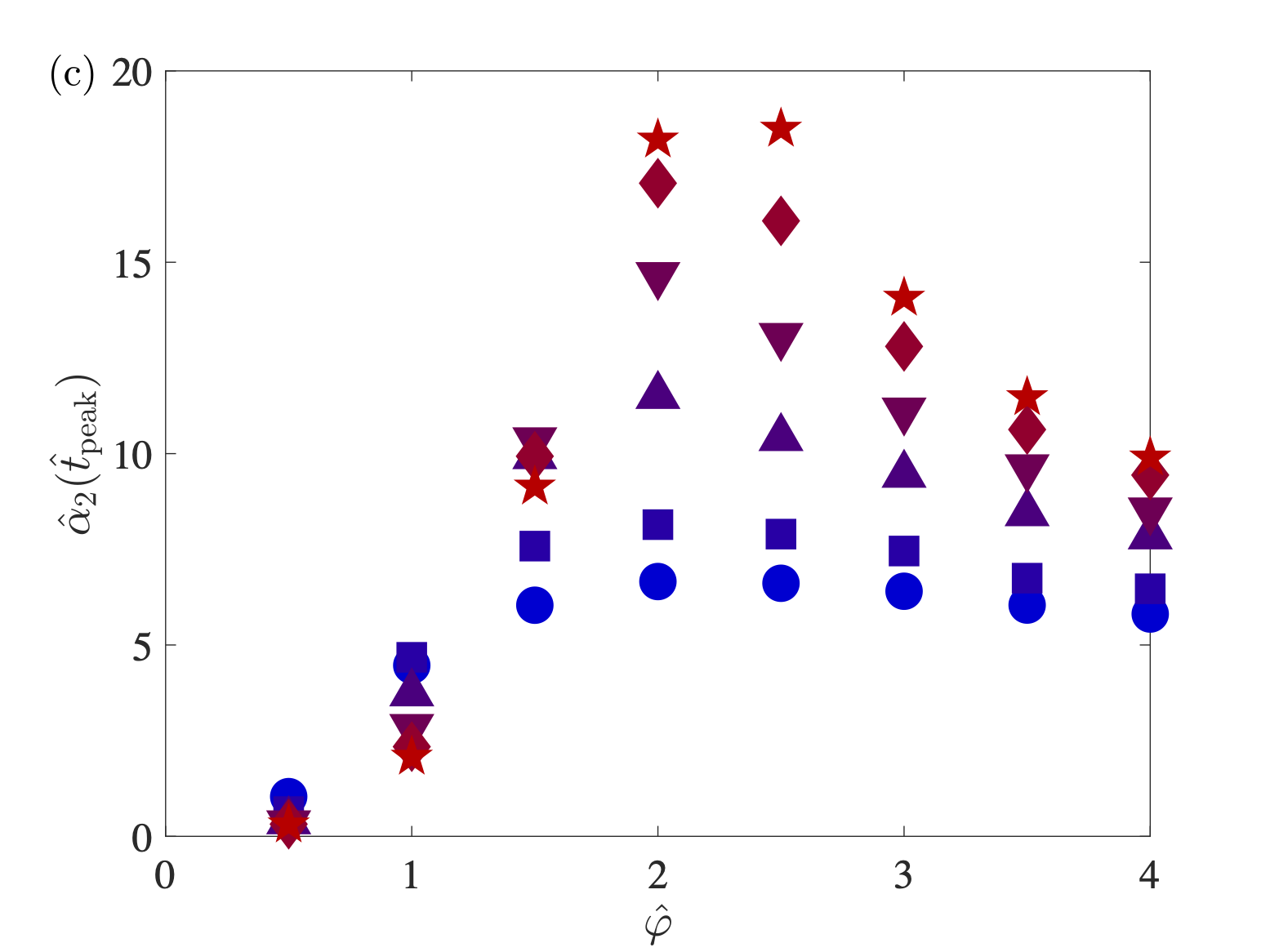}
\includegraphics[scale=0.45,trim={0.8cm 0 0.1cm 0.2cm},clip]{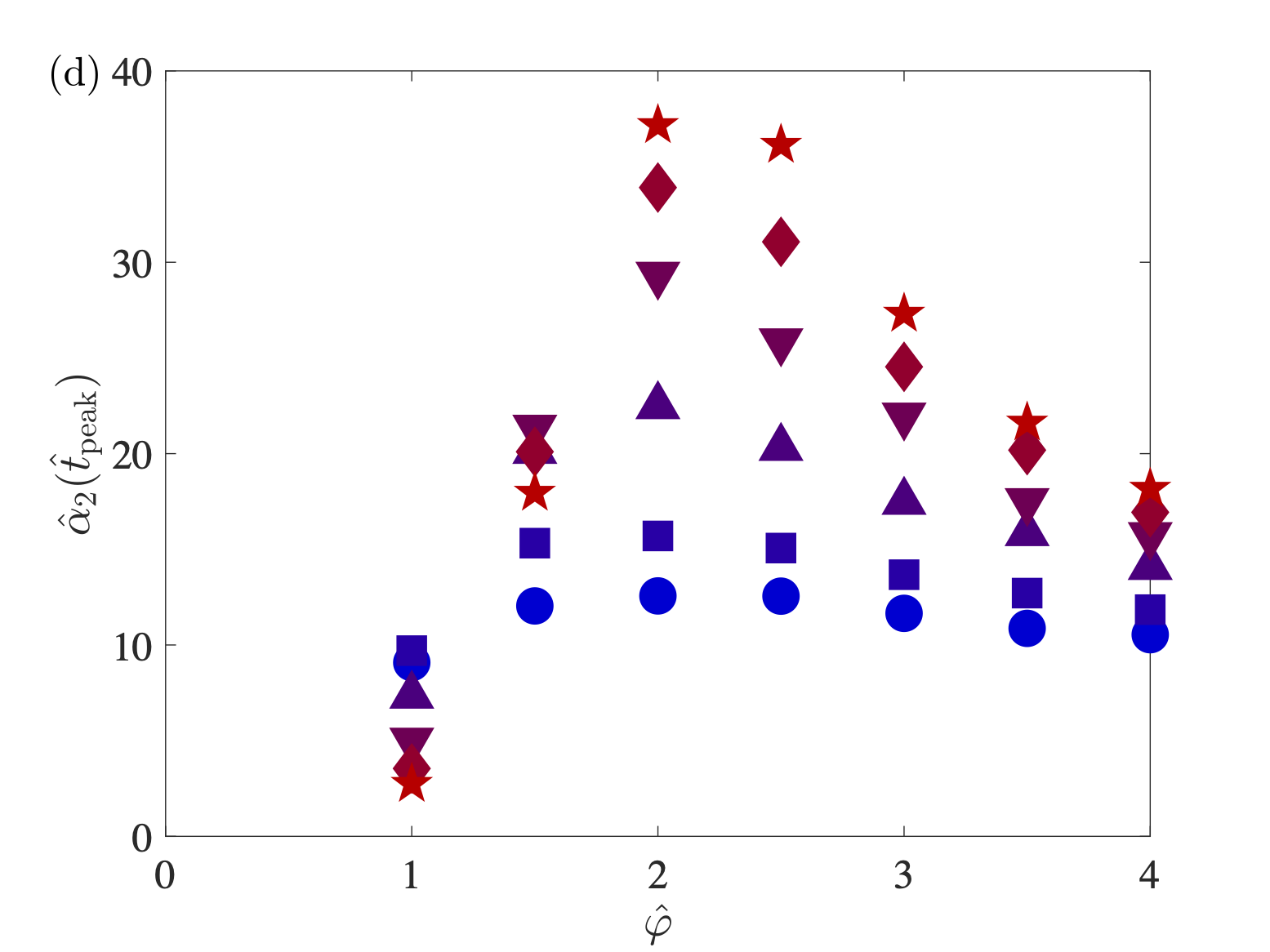}
\caption{The heights of the short-time peak in $\hat{\alpha}_2$ of the ABP-RLG with $\widehat{\mathrm{Pe}}=$ (a) $3$, (b) $5$, (c) $10$ and (d) $20$, and for various $d$.}
\label{fig:si_a2_peak1}
\end{figure}

Fig.~\ref{fig:si_a2_peak1} shows the height of the first peak for various $\widehat{\mathrm{Pe}}$, and leads to data reported in the inset of Fig.~4. The results show similar trends with $\hat{\varphi}$ and $d$; panel (c) is provided in the main text.

\printbibliography